# Assessing Semantic Quality of Web Directory Structure

Marko Horvat, Gordan Gledec and Nikola Bogunović

Faculty of Electrical Engineering and Computing, University of Zagreb
Unska 3, HR-10000 Zagreb, Croatia
E-mail: {Marko.Horvat2, Gordan.Gledec, Nikola.Bogunovic}@fer.hr

**Abstract.** The administration of a Web directory content and associated structure is a labor intensive task performed by human domain experts. Because of that there always exists a realistic risk of the structure becoming unbalanced, uneven and difficult to use to all except for a few users proficient in a particular Web directory. These problems emphasize the importance of generic and objective measures of Web directories structure quality. In this paper we demonstrate how to formally merge Web directories into the Semantic Web vision. We introduce a set of objective criterions for evaluation of a Web directory's structure quality. Some criteria functions are based on heuristics while others require the application of ontologies.

## 1 Introduction

The Semantic web vision and related spectrum of technologies have enjoyed rapid development during the last eight years. The initial paper by Tim Berners-Lee [1] introduced the notion of universally described semantics of information and services on the Web. The vision of a Web as a shared common medium for data, information and knowledge exchange, and collaboration, fostered a wealth of development and research. The idea itself was simple but appropriately far reaching. The Semantic web brought the power of managed expressivity provided by ontologies to the World Wide Web (WWW) [2]. Today the research in Semantic web application is not largely focused on the problem of ontologically-based Web directories [3][4][5]. Furthermore, as yet a lot of the effort is unfinished and more systems are in the phase of research and development (R&D) than in everyday production [6].

However, Web directories have simple hierarchical structures which are commonplace and effective for data storage importance and classification. This makes them important applications for data storage or classification, and motivates research in the assessment of their semantic qualities.

The remainder of the paper is organized as follows; the next chapter describes the categories and the structure of Web directories. Mutual associations between the Semantic Web and Web directories, as well as the semantic dimension of categories, are all presented in the third chapter. Quality measures of Web directories are discussed in the fourth chapter. Related publications and our conclusion with outlook for future work are presented at the end of the paper.

## 2    Categories and the structure of Web directories

In order to explain how Web directories can be positioned within the Semantic web vision it is first necessary to formally define all constituent elements and how they organize to make up a Web directory, and secondly to add semantic annotations to these building elements. A web directory, web catalog or link directory as it is also called, is a structured and hierarchically arranged collection of links to other web sites. Web directories are divided into categories and subcategories with a single top category, often called the root category, or just the root. Each category can have a provisional number of subcategories with each subcategory further subsuming any number of other subcategories, and so on. Furthermore, every category has a unique name and an accompanying Uniform Resource Locator (URL), and can also carry other associated information.

Each category of a Web directory contains a set of links to various sites on the WWW, and a set of links to other categories within the web directory. This basic trait is the most important feature of a Web directory.

Each Web directory has a start page, i.e. a home page, which represents its root category, and every other category of a Web directory has its own adjoined web page. The start page displays subcategories that belong to the root. By following a link to a subcategory, user opens that category's page and browses through its links and subcategories. This process continues until the user finds a link to a web resource that s/he is looking for. In essence, the user can be described as an intelligent agent that traverses the structure of a Web directory looking for specific information.

Since Web directories are always rooted and the order of categories is strictly maintained, it is possible to assign level numbers to categories. The subcategories of the root are the $2^{nd}$ level categories, and in turn their subcategories are the $3^{rd}$ level categories, and so on. Maximum level of a Web directory is called depth.

Each category, except the root, has one category above it, which is called its parent. The categories below a certain category (i.e. with a greater level number than the category) are called its children, while categories on the same level as a node are called its siblings. Categories with no children are called terminal categories, and a category with at least one child is sometimes called nonterminal category. Associations between categories are arbitrary, but there must be at least one path between any pair of categories. Disjoint sets of categories are not allowed, as well as parallel links and self-loops. Each nonterminal category must have links to all its children, but can also have links to other categories in the Web directory which are semantically similar, or otherwise analogous to the category.

We will formally designate with ***C*** the set of all categories in a Web directory, and ***R*** will be the set of all Web resources in a Web directory. One category with unique identification number *n* is denoted $c_n$. Category has its own characteristic URL *url*.

The category $c_n$ must be a member of **C**. $C_n$ is a subset of **C** that belongs to the category $c_n$, and $R_n$ the subset of **R** with Web resources that belong to the category $c_n$. In order to be more informative, the categories can also be written as $c_n^l$ with their member level $l$, where $l$ is a natural number smaller than or equal to the depth of a Web directory $L$. Therefore, category is a tuple $c_n = \{n, l, url, C_n, R_n\}$ and can be schematically annotated as in the figure below.

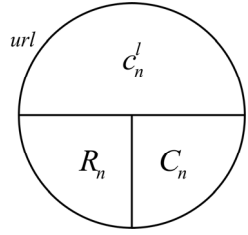

**Fig. 1.** Schematic representation of a single category.

We can define a Web directory $wd$ to be an element of the WWW. With **C** and **R** being members of $wd$ the algebraic definitions of the elements of a Web directory and their mutual relationships are

$$\begin{aligned} C_n &\in c_n \subset \mathbf{C} \\ R_n &\in c_n \subset \mathbf{R} \\ l &\in [1, L] \in \mathbf{N} \\ \mathbf{C} &\in wd \subset \text{WWW} \\ \mathbf{R} &\in wd \subset \text{WWW} \end{aligned} \quad (1)$$

The set of all children categories to $c_n^l$ is $C_n$ while the set of all children one level below is $C_n^{l-1}$ or $C_n^{-1}$, two levels below $C_n^{l-2}$ or $C_n^{-2}$, etc. As can be seen in (1) category is also a Web resource ($c_n \subset \mathbf{R}$), as it should be expected since it has unique URL and carries specific information. Furthermore, Web directory itself also becomes a tuple $wd = \{\mathbf{C}, \mathbf{R}\}$.

Mathematically speaking, Web directories are simple rooted graphs [7]. In this formal respect, categories represent vertexes and connections represent vertices. The path between two vertexes is called the arc, edge or link, and when there is an edge connecting the two vertices, we say that the vertices are adjacent to one another and that the edge is incident on both vertices. The degree of a vertex is equivalent to the number of edges incident on it.

Using the described formalisms, the schema of a simple Web directory with 6 categories distributed in 4 levels, with parent-child associations and two specific links $c_6 \to c_5$ and $c_3 \to c_2$ could be depicted with Fig. 2.

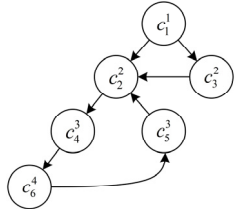

**Fig. 2.** Schematic representation of a Web directory.

However, the structure of a Web category (Fig. 3) cannot be described as a tree because more than one path can connect any of its two categories: apart from paths which connect parent/child categories, they can be associated with *ad hoc* cross-links.

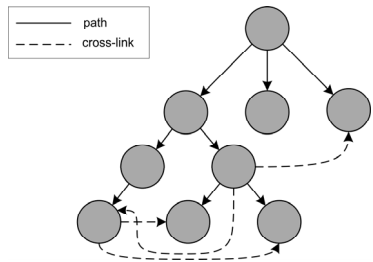

**Fig. 3.** Realistic Web directory with possible multiple paths between two categories.

If, for the sake of discussion, all categories of a Web directory except the root had paths only to its children such structure would constitute a rooted tree, as in Fig. 4.

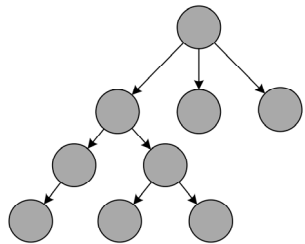

**Fig. 4.** Idealistic Web directory with only one path between any two categories.

Sometimes the order of categories appearance is relevant, e.g. the position of links within a category's Web page is prioritized, and in that case we are talking about ordered and rooted simple graphs as an algorithmic definition of a Web directory.

Although the categorization of a Web directory should be defined by a standard and unchanging policy, this is frequently not the case. Web directories often allow site owners to directly submit their site for inclusion, even suggest an appropriate category for the site, and have editors review the submissions. The editors must approve the submission and decide in which category to put the link in. However, rules that influence the editors' decision are not completely objective and are thus difficult to implement unambiguously. Sometimes a site will fall in two or even more categories, or require a new category. Defining a new category is a very sensitive task because it has to adequately represent a number of sites, avoid interfering with domains of other categories, and at the same time the width and depth of the entire directory's structure has to be balanced. A Web directory with elaborate structure at one end and sparse and shallow at the other is confusing for users and difficult to find quality information in. Furthermore, after several sites have been added to a directory it may become apparent that an entirely new categorization could better represent the directory's content. In this case a part of directory's structure or even all of its levels have to be rearranged which is again time and labor consuming task.

## 3    The Semantic Web and Web directories

At the moment of writing, the resources on the WWW are primarily designed for human and not machine use [8]. To rephrase it, the declarative and procedural knowledge currently offered by various Web sources is shaped in a way that better suites humans and not machines. The vision of the Semantic Web is directly aimed at solving this dichotomy by introducing self-describing documents that carry data and the accompanying metadata together, and thus organize and interconnect available information so it also becomes processable by computer applications [9].

The structure of a Web directory is basically a subjective construction. It depends on human comprehension and the policy taken by the Web directory's administrator, or even on the users that submit sites to the directory. It is important to note that not all Web directories, or even all segments of a Web directory, have the same editorial policy. Clearly, for the sake of a Web directory's informative clarity and usability, the semantic distance between any two categories should be approximately constant, and not dramatically vary from one category to the next. Also, the key for the selection of concepts that represent categories should remain uniform throughout the directory's structure. The only parameters that should be used to judge the quality of a directory are its informative value and usability, to humans and machines equally. In the fifth chapter we will propose several numerical parameters that objectively measure the worth of a directory.

Let's assume that we have at a disposition function *sem* that takes a resource $r_i \in \boldsymbol{R}$ and from its semantic content builds an ontology $o_i \in \boldsymbol{O}$ where $\boldsymbol{R}$ and $\boldsymbol{O}$ are sets of all resources and ontologies, respectively.

$$sem : \boldsymbol{R} \rightarrow \boldsymbol{O} \tag{2}$$

The function *sem* builds an ontology from a resource. In slightly different terms, it creates a solid representation of an abstract property. This property can be described as informal and explicit on the semantic continuum scale [10] and its technical realization is strictly formal. Operations of the function *sem* can be performed by a computer system or a domain expert, in which case we talk about automatic or manual ontology construction, respectively. The necessary mathematical assumption on *sem* is it has well-defined addition and subtraction operators in **R** and **O**

$$\oplus : R \times R \to R \quad (3)$$
$$\odot : R \times R \to R$$
$$\hat{+} : O \times O \to O$$
$$\hat{-} : O \times O \to O$$

This allows application of union operator across these two sets and concatenation of individual resources and ontologies, as well as determining their respective differences

$$sem(r_1 \oplus r_2) = sem(r_1) \hat{+} sem(r_2) \quad (4)$$
$$sem(r_1 \odot r_2) = sem(r_1) \hat{-} sem(r_2)$$

Also, we should define a modulo operator $|\bullet|$ on **O** as $|\bullet| : O \times O \to O$.

The *semantic content of a category* can be defined in three ways: *i)* by its Web resources, *ii)* from its subsumed categories, *iii)* as a constant.

By the first definition, semantic content of a category $c_i$ within a Web directory *wd* derives from the semantic content of all its Web resources $r_{ij}$ where $r_{ij} \in R_i \in c_i$ as

$$sem(c_i) = \hat{+}_{r_{ij} \in R_i} sem(r_{ij}) \quad (5)$$

According to the second definition, the semantic content of $c_i$ can also equal the aggregation of the semantic content of its children categories $c_j \in C_i^{-1} \in c_i$

$$sem(c_i) = \hat{+}_{c_j \in C_i^{-1}} sem(c_j) \quad (6)$$

Finally, if $c_i$ has no resources $(R_i = \varnothing)$ and subcategories $(C_i = \varnothing)$ it is assumed that the semantic content of $c_i$ is defined by a constant $const_i$ as

$$sem(c_i) = const_i : R_i = \varnothing, C_i = \varnothing \quad (7)$$

Reasoning behind such threefold definition is that the meaning of categories is conformed to the directory's editorial policy. If a category is empty and no resources have been added, it will still have some member semantics attached by the Web directory administrator.

The structure of directory *wd* is *ideal* if for non-empty **R** and **C**

$$\left| \hat{+}_{r_{ij} \in R_i} sem(r_{ij}) \stackrel{\_}{\_} \hat{+}_{c_j \in C_i} sem(c_j) \right| = \emptyset \qquad (8)$$
$$\forall c_i \in wd, R_i \in c_i, C_i \in c_i$$

That is, the structure of directory *wd* can be considered perfect if and only if for each category $c_i \in wd$ the semantic content of its Web resources $R_i \in c_i$ and subsumed categories $C_i \in c_i$ are equal.

Pragmatically, we can define a neighborhood $\varepsilon$ within **O** and say that the structure of directory *wd* is *realistically ideal* if

$$\left| \hat{+}_{r_{ij} \in R_i} sem(r_{ij}) \stackrel{\_}{\_} \hat{+}_{c_j \in C_i} sem(c_j) \right| \leq \varepsilon \qquad (9)$$
$$\forall c_i \in wd, R_i \in c_i, C_i \in c_i$$

The existence of the function *sem*, with the described properties, is fundamental and indivertible in the ontology-based construction of Web directories.

## 4    Semantic quality measures

During or after Web directory's construction it is highly desirable to establish some measures of value of the accomplished process. The criterion functions that will provide these measurements should be objective and universal. Benefits of such measures would be twofold: *i*) they could provide a matching framework between Web directories, and *ii*) they could be used to assess the semantic structure quality of individual Web directories. In other words, by using them structures of any two Web directories could be objectively compared and the criterions could point to potential semantic deficiencies in a directory. Information retrieval in Web directories can be executed either through searching or browsing scenarios. Because of the sheer size of data available on the Web, searching is the dominant retrieval scenario. Several performance measures for evaluation of searching scenarios have already been proposed, such as precision, recall, fall-out and F-measure. However, information seeking by browsing scenarios is interesting in reduced information collections like blogs, RSS feeds, social networks [11], but also individual directory categories. Since information in Web directories may be browsed by intelligent agents as well as human users, the establishment of parameters for objective measurement of Web directory's structure and content is of a significant importance for determining its usability, semantic quality and subsequently other intrinsic characteristics.

We have identified three parameters that can be used to objectively assess the semantic quality of a Web directory. The parameters are:

1. Path ratio
2. Maximum revisit
3. Distance decrease progression

All parameters require observation of user's actions, i.e. browsing pattern of a person or an intelligent agent using the directory. We will assume that the browsing scenario starts at the root category although this is not strictly necessary (nor is often the case in real-world use). The parameters are calculated based on observation of an action of a single user. Each observation represents one browsing session for a specific resource contained within the directory. After the parameters of individual observations are collected they may be statistically processed and aggregated. This data can then cumulatively represent relevant trends and features in the actions of any number of directory's users.

*Path ratio* (PR) is calculated as a proportion between the minimum number of categories between the root and the category with the required Web resource, and the number of categories the user traverses while browsing. Therefore, when browsing for a resource $r$ in a Web directory $wd$ the browse $b(r,wd)$ with the length $|b(r,wd)|$ parameter PR is defined as

$$PR(b) = 1 - \frac{\min|b(r,wd)|}{|b(r,wd)|}, PR(b) \in [0,1\rangle \qquad (10)$$

The rationale behind this parameter is that in the case of the optimal, or direct, browse $b^*$ the user will achieve the shortest path between the root and the category with the resource browsed for. In this case $PR(b')=0$. In a suboptimal, or indirect, browse $b'$ user will traverse at least one category more and $PR(b')>0$. This is explained in the next figure that illustrates a browsing pattern staring at category $c_1^1$ and ending at $c_9^4$.

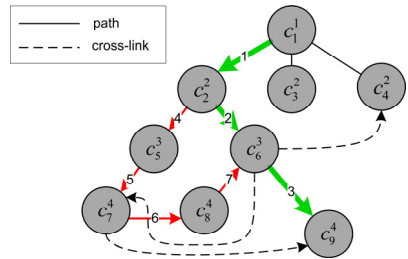

**Fig. 5.** Optimal (direct) and suboptimal (indirect) browse paths in calculating parameter PR.

Browse $b_1$ with the path 1→2→3 is optimal because it traces the shortest and the most direct path between the start and the end category so that $PR(b_1)=0$. While the browse 1→4→5→6→7→3 will also lead to the end resource, it is suboptimal since its length is greater than that of the optimal browse (6 > 3), thus PR($b_2$) = 1/2.

*Maximum revisit* (MR) or *maximum category revisit* is a parameter that describes the maximum number of repeated visits to any category while browsing for one resource. Because Web directories are simple rooted graphs with at least one path between any two nodes, there is never a need to visit the same category twice while browsing for a resource. Therefore, MR specifies the level of wander or loitering in a Web directory's structure while browsing (Fig. 9).

The best possible browse $b$ for a resource $r$ in a Web directory $wd$ has

$$\mathrm{MR}\big(b(r, wd)\big) = 0 \qquad (11)$$

indicating no category revisit, where MR($b$) can be any natural whole number including zero $\mathrm{MR}(b) = 0, 1, 2, ..., n, n+1 \in N \cup \{0\}$.

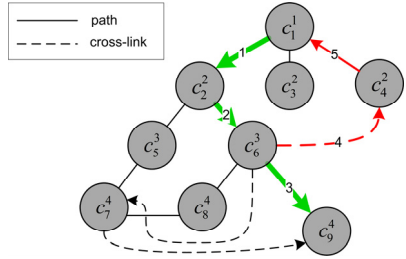

**Fig. 6.** Optimal and suboptimal browsing paths with revisits in calculating parameter MR.

In the Fig 6., browse $b_1$ with the path 1→2→3 starting in $c_1^1$ and finishing with $c_9^4$ has MR($b_1$) = 0. However, due to the configuration of the directory it is possible to needlessly revisit some or even all categories. This is illustrated in the browse $s_2$ with the path 1→2→4→5→1→2→3 which gives MR($b_2$) = 1. Since MR($b_1$) < MR($b_2$) browse $b_1$ is a better then $b_2$.

*Distance decrease progression* (DDP) is an ontology-based parameter. It describes the gradient of semantic convergence toward the resource during one browse. As the user browses categories looking for a particular resource, each category s/he visits should be progressively ontologically closer to the resource. If this is not the case, than either he is loitering or the directory does not have the optimal structure. Parameter DPP($s$) can be defined as a series

$$\mathrm{DPP}(b) = \sum_{i=1}^{n-1} dist(c_i, c_T) - dist(c_{i+1}, c_T) \qquad (12)$$

where $c_T$ is the target category containing the resource the user is looking for, $c_i$ is any category being browsed and $n$ is the length, i.e. number of steps, of the browse $b$. It is also necessary to apply a similarity measure $sim : C^2 \to [0,1]$ between the two categories $c_1, c_2 \in C$ and a distance function $dist(c_1, c_2) = 1 / sim(c_1, c_2)$ as defined in [12][13]. If the sequence of partial sums $\{s_1, s_2, ..., s_n, s_{n+1}, ...\}$ converges, than the series is also convergent, where

$$s_n = \sum_{k=1}^{m} dist(c_k, c_T) - dist(c_{k+1}, c_T) \tag{13}$$

The search $b'$ is optimal if DPP($b'$) converges to 0.

All three parameters described here should be used in conjunction with each other in order to cumulatively describe this important design feature of Web directories.

Node distribution in some Web directories, at a certain level in their structures, does not necessarily have to follow concept semantics partition or this process can be somehow affected and skewed. Examples of this are content division according to date, contributors' names or alphabet, e.g. having node "A" for subnodes with "Apples" content, "B" for "Bananas", "C" for "Citrus", etc. These nodes would have more in common with a concept "Fruit" that with "Alphabet Letter". Subsequently, mutual semantic distance of such nodes would be great and incompatible with the directory's partition. In order to overcome this problem in calculation of the semantic quality parameters one has to simply ignore semantic value of these nodes at a level *l* and directly link nodes in levels *l-1* and *l+1*. By doing this monotone semantic difference between nodes is restored.

Every Web directory should have an easily understandable semantic schema that is reflected in a directory's structure so it becomes self-explanatory which category to browse in order to iteratively and progressively approach the required resource. This issue is closely correlated to the Web usability of directories. However, due to diverse quality of data sources available on the Web it is not easy to construct a directory with an ideal path ratio, maximum revisit and distance decrease progression values. Further planned experiments should provide more information on the everyday applicability of the parameters proposed here.

## 5     Related work

All previous work regarding coupling of Web directories with ontologies and the Semantic web paradigms have been directed at using Web directories, their data and structure, to extract information from WWW with the goal of document classification and ontology learning. In this paper we presented an exactly opposite approach – using available knowledge to construct a Web directory itself.

The paper by Kavalec [4] which described a mechanism for extraction of information on products and services from the source code of public web pages was especially useful in our work. Papers by Mladenić [5], Li [14] and Brin [15] were also helpful.

We would particularly like to emphasize the work by a research group at FER which introduced ontologies in the search mechanism of the Croatian Web directory, and thus successfully resolved problems of low recall, high recall and low precision and vocabulary mismatch [16]. The Croatian Web directory (http://www.hr/) [17] was founded in February 1994 and its purpose has been to promote and maintain the network information services through the "national WWW homepage" and enable easy navigation in Croatian cyberspace using hierarchically and thematically organized directory of WWW services. At the moment of writing the directory contains 25,185 Web resources listed in 753 categories.

## 6    Conclusion

Web directories are commonplace method for structuring semantically heterogeneous resources. The form of simple rooted graphs is well-suited for information storage and representation in Web and Desktop environments, and in numerous applications ranging from directory trees, bookmarks and generic menus to tables of content. Also, the aspect of social collaboration is very important since networking and the Web enable instant publication and usage of data, ideally within groups of trusted users with the same areas of interest. All this only emphasizes the importance of successful construction, management, information extraction and reuse of all simple rooted graphs data structures and Web directories in particularly.

We would like to advise caution in using publicly available Web directories to learn new ontologies. Structures of Web directories are often biased and influenced by the contributors of resources. Administration of a large directory is an overwhelming task prone to errors. Therefore, it may be better to construct ontologies from smaller directories or from directories with rigid administrative policies. The former directories are more numerous than the latter, but they will also offer less information and in a more specialized area.

The primary goal of this paper was to put forward a series of objective criteria functions for evaluating the quality of Web directories. Two criteria are based in heuristics while the third calls for introduction of ontologies, which is possible only if Web directories are placed in the context of the Semantic web vision.

In the future we would like to apply the presented measures to a real Web directory and through a set of measurements obtain experimental results with relevant statistics about its structure quality. In this we plan to use the Croatian Web directory and its domain "Tourism" as a suitable test category. Also, we shall combine this work with our other efforts to develop ontology-based software for automated construction of Web directories [18]. In this respect the semantic quality measures would be used as control parameters in an iterative process of constructing and refining the Web directory's structure.

**Acknowledgements.** We would wish to thank Siniša Popović at the Laboratory for Interactive Simulation Systems (http://liss.esa.fer.hr/), Faculty of Electrical Engineering and Computing for his kind assistance in defining the algebra used in the paper.